\documentclass[iop,apj]{emulateapj}
\usepackage{amsmath,amssymb,amstext}
\usepackage[breaklinks,colorlinks,citecolor=blue,linkcolor=magenta]{hyperref} 

\usepackage[all]{hypcap}
\usepackage{graphicx}
\usepackage{subfigure}
\usepackage{booktabs}
\usepackage[flushleft]{threeparttable}
\usepackage{float}

\begin{document}
\title{SARAS 2 constraints on global 21-cm signals from the Epoch of Reionization}
\author{Saurabh Singh$^{1,\dag}$}
\author{Ravi Subrahmanyan$^1$}
\author{N. Udaya Shankar$^1$}
\author{Mayuri Sathyanarayana Rao$^1$}
\author{Anastasia Fialkov$^2$}
\author{Aviad Cohen$^3$}
\author{Rennan Barkana$^3$}
\author{B.S. Girish$^1$}
\author{A. Raghunathan$^1$}
\author{R. Somashekar$^1$}
\author{K.S. Srivani$^1$}
\thanks{$^\dag$Joint Astronomy Program, Indian Institute of Science, Bangalore 560012, India }
\affil{{\small $^{1}$Raman Research Institute, C V Raman Avenue, Sadashivanagar, Bangalore 560080, India}}
\affil{{\small $^{2}$Harvard-Smithsonian Center for Astrophysics, Institute for Theory and Computation, 60 Garden Street, Cambridge, MA 02138, USA}}
\affil{{\small $^{3}$Raymond and Beverly Sackler School of Physics and Astronomy, Tel Aviv University, Tel Aviv 69978, Israel}}
\email{Email of corresponding author: saurabhs@rri.res.in}

\begin{abstract}

Spectral distortions in the cosmic microwave background over the 40--200~MHz band are imprinted by neutral hydrogen in the intergalactic medium prior to the end of reionization. This signal, produced in the redshift range $z = 6-34$ at the rest frame wavelength of 21 cm, has not been detected yet; and poor understanding of high redshift astrophysics results in a large uncertainty in the expected spectrum.  The SARAS~2 radiometer was purposely designed to detect the sky-averaged 21-cm signal. The instrument, deployed at the Timbaktu Collective (Southern India) in April--June 2017, collected 63~hr of science data, which were examined for the presence of the cosmological 21-cm signal. In our previous work the first-light data from SARAS~2 radiometer were analyzed with Bayesian likelihood-ratio tests using  $264$ plausible astrophysical scenarios. In this paper we re-examine the data using an improved analysis based on the frequentist approach and forward modeling. We show that SARAS~2 data rejects 27 models, out of which 25 are rejected at a significance $>5\sigma$. All the rejected models share the scenario of inefficient heating of the primordial gas by the first population of X-ray sources along with rapid reionization. 

\end{abstract}


\section{Introduction}
The Universe at the epochs of Cosmic Dawn (CD) and Reionization (EoR) is poorly constrained by observations, which results in a large scatter in theoretical predictions for galaxy and star formation. One of the most powerful potential probes of these eras is the rest-frame 21-cm signal of neutral hydrogen (HI) produced by the intergalactic medium (IGM) prior to the end of the EoR at $z\sim 6$. 
The intensity of this signal is tied to the star formation history as well as to the ionization and thermal histories of the IGM \citep{Barkana:2016}. Hence, its measurement will  bracket astrophysical properties of the first UV and X-ray sources including the ionizing efficiency of first stars and quasars, luminosity and spectra of the  first population of black holes, and properties of dark matter particles \cite[]{2010PhRvD..82b3006P, 2013ApJ...777..118M, 2017MNRAS.472.1915C, 0004-637X-813-1-11, 2014Natur.506..197F, 2006MNRAS.371..867F, 2017arXiv171002530M, Sitwell:2014, Evoli:2014}. At present these properties are poorly understood allowing for a large variety of plausible 21-cm spectra \citep{2017MNRAS.472.1915C}. 

The main feature of the sky-averaged (a.k.a. global) 21-cm spectrum observed against the cosmic microwave background (CMB) is the deep absorption trough which traces the adiabatic cooling of the IGM and its subsequent heating by the first X-ray sources (X-ray binaries and black holes). The thermal history is imprinted in the 21-cm signal owing to the Wouthuysen-Field effect \citep{Wouthuysen:1952, Field:1958}: the 21-cm transition is coupled to the temperature of the gas by the stellar Ly$\alpha$ photons. The strength of the coupling  depends on the intensity of the Ly$\alpha$ background and is correlated to  the process of star formation itself. When the population of X-ray sources builds up producing sufficient amount of photons with energy in the $\sim 0.1-3$ keV range, temperature of the IGM rises leading to a reduction in the 21-cm intensity and shaping the absorption trough. Considering $\sim 200$ different plausible astrophysical scenarios \citet{2017MNRAS.472.1915C} show that, owing to the uncertainty in the high-redshift astrophysics,  the depth of the absorption trough can vary between --25 and --240 mK and its central frequency can be anywhere between $40<\nu <120$ MHz  (corresponding to $z\sim 11-34$). Localization of this feature will directly constrain the intensity of the  Ly$\alpha$ background and the cosmic heating rate. 

At lower redshifts $z\sim 6-11$ (higher frequencies  $\sim 120-200$ MHz) reionization by stars and quasars is ongoing, and the intensity of the 21-cm signal decreases owing to the lesser fraction of HI in the IGM. As more free parameters are added to the modeling (e.g., ionizing efficiency of sources and mean free path of the ionizing photons), the expected signal is even less constrained. In particular, its shape  depends on the balance between the heating and ionization rates: if heating occurs faster than ionization, the signal will be seen in emission during the EoR, otherwise it will be seen in absorption at any epoch. If it is present, the emission feature can be as strong as 32 mK peaking between 80--160 MHz. Detecting the EoR signal will allow to constrain X-ray heating efficiency together with the ionization efficiency of  sources \citep{2017MNRAS.472.1915C}; it will also measure the CMB optical depth $\tau$ at much higher precision than what can be done with the CMB \citep{Liu:2016, Fialkov:2016}.

Ongoing experiments that target detection of the global 21-cm signal from CD and EoR are plagued by orders of magnitude stronger Galactic and extragalactic foregrounds \cite[]{1999A&A...345..380S,2017ApJ...840...33S}. These foregrounds couple to the radiometer system through its frequency dependent transfer function and can potentially confuse a detection of the relatively faint cosmological 21-cm signal.  Additional challenges include modeling the internal additives from within the receiver system, which are often difficult to calibrate, and excision of terrestrial Radio Frequency Interference (RFI).  All these demand stringent requirements on the antenna and receiver design, clever calibration strategies and innovative data analysis methods \cite[]{2017arXiv171001101S}. 

Despite the challenges, pioneering experiments have attained sensitivity levels at which plausible scenarios of reionization are being ruled out. The first constraint on EoR from global 21-cm experiments came from the Experiment to Detect the Global EoR Signature (EDGES) high band antenna covering 90--190 MHz frequency range, which ruled out rapid reionization with $\Delta z < 0.06$ at the  $95\%$ confidence level \cite[]{2010Natur.468..796B}. \citet{2016MNRAS.461.2847B} used an  outrigger Large Aperture Experiment to Detect the Dark Ages (LEDA) antenna to measure the spectrum at lower frequencies, 50--100 MHz. This measurement constrained  the  amplitude of the absorption trough to be less than $890~\rm mK$  for a Gaussian-shaped absorption with width greater than $6.5 \rm~MHz$ at the 95\% confidence level.  Constraints on the redshift interval, $\Delta z$, over which reionization occurred have significantly improved with the recent high-band data from EDGES \cite[]{2017ApJ...847...64M}. The constraint depends on the assumptions for the thermal state of the IGM during the EoR: for heated IGM models, the duration shorter than $\Delta z \approx 1$ with EoR happening at $z\approx 8.5$ is rejected with $95\%$ confidence; whereas for cold IGM scenarios, $\Delta z \lesssim 2$ is rejected over most of the plausible redshift range for the EoR. All the analyses mentioned above were carried out adopting simple functions to mimic the cosmological signal:  a $\\tanh$ form was used to imitate the variation in ionization fraction with frequency, and the absorption trough was modeled as a Gaussian. 

Realistic global 21-cm signals \citep[part of which were published in][]{2017MNRAS.472.1915C} were used for the first time in the  analysis of the first-light data of Shaped Antenna measurement of the background RAdio Spectrum~2 (SARAS~2) radiometer \cite[]{2041-8205-845-2-L12}. The spectra are outputs of  a self-consistent 4-D (3 spatial dimensions + time) large-scale simulation of the high redshift universe \citep[e.g.,][]{Visbal:2012, Fialkov:2014}. In this simulation X-ray and UV photons emitted by a realistic non-uniform and time-dependent population of sources are propagated accounting for time delay and cosmological redshift. These photons heat and ionize the initially cold and neutral IGM which produces the 21-cm signal.  Using Bayesian likelihood-ratio tests the SARAS~2 data were shown to disfavor 9 out of 264 different astrophysical scenarios with $1\sigma$ confidence over the rejected set. All these models share late IGM heating along with rapid reionization \cite[]{2041-8205-845-2-L12}. 

In this paper we employ improved statistical techniques to analyze the same data of SARAS~2 and use the same set of astrophysical models as in \citet{2041-8205-845-2-L12}. We adopt the frequentist approach of \cite{2017ApJ...847...64M}, including forward modeling, and revisit the likelihoods for each one of the cosmological signals. The paper is organized as follows. Section~\ref{sec:desc} summarises the SARAS~2 system and the observations.  Section~\ref{sec:ana_str} outlines the data analysis  method.  In Section~\ref{sec:cons} we discuss astrophysical constraints. We discuss limitations of the analysis methods in  Section~\ref{sec:cav}.  We conclude in Section~\ref{sec:conclusion}.  

\section{SARAS~2: a description of the radiometer and observations}
\label{sec:desc}
SARAS~2 is a precision radiometer, custom designed to detect global 21-cm signal from CD \& EoR, covering the band 40--200 MHz which corresponds to redshift range $z \sim$ 6--34. SARAS~2 has been designed to have (i) a telescope beam that is frequency independent so that structure in the foreground sky brightness does not result in any spectral shapes in the response, and (ii) a receiver transfer function and internal systematics---both multiplicative and additive---that are spectrally smooth so as to allow a separation of foregrounds and systematics from the predicted global cosmological 21-cm signals.  A detailed description of the system design and calibration scheme of SARAS~2 was presented in \cite{2017arXiv171001101S}. 

The system was deployed at a relatively radio quiet site at the Timbaktu Collective in Southern India during 2017 April--June.  The data were processed to reject RFI, calibrate the receiver gain and bandpass, and the data along with the GMOSS model \cite[]{2017AJ....153...26S} for the radio sky were used to derive the total efficiency of the radiometer.  A total of 63~hr of useful night time data were obtained over the frequency band of 110--200~MHz. Data residuals, after modeling for foregrounds and internal systematics, yielded spectra with resolution 122~kHz and root-mean-square (RMS) noise of 11~mK, consistent with expectations from the radiometer system temperature, observing time etc.  First results of the  first-light data from SARAS~2 were presented previously in \cite{2041-8205-845-2-L12}. 

\section{Signal Extraction: A frequentist approach}
\label{sec:ana_str}

In our first paper \citep{2041-8205-845-2-L12} we have used Bayesian likelihood-ratio tests to verify whether or not each one of the theoretical signals is consistent with the first-light data. Here we use the same data and the same set of models, but a different statistical approach. 

\subsection{Foreground Modeling}

The observed data consists of the cosmological and foreground signals, propagated through the SARAS 2 system,  plus the internal systematics generated by the instrument. Both the foreground and the systematics are modeled using polynomials over an optimal frequency band (as described in subsection \ref{subsec:sens}). The total contribution of foregrounds and systematics is thus $F(\nu) = \sum_{i=0}^N c_i \nu^i $, where  $c_i$ are the $(N+1)$ coefficients of the polynomial. 
\\
\subsection{Signal Propagation}

\begin{figure}[htbp]
\begin{center}
\includegraphics[scale=0.45]{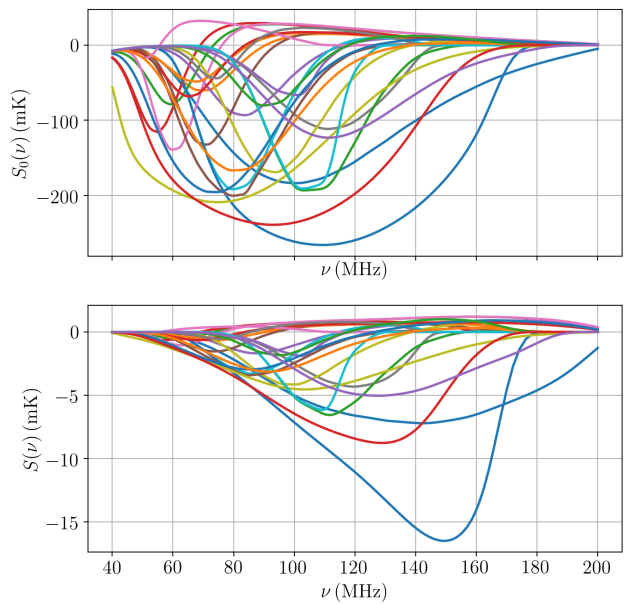}
\caption{Top: A representative set of 25 input global 21-cm spectra $S_0(\nu)$ as a function of frequency in mK units \citep{2017MNRAS.472.1915C}. Bottom: propagated spectra, $S(\nu)$.}
\label{fig:eff}    
\end{center}   
\end{figure}

The cosmological 21-cm signal propagated through the SARAS~2 system, $S(\nu)$, is related to the input cosmological signal, $S_0(\nu)$, by the total efficiency $\eta_t(\nu)$ of the SARAS~2 monopole antenna; {\it i.e.}, $S(\nu) = \eta_t(\nu) \times S_0(\nu)$. To model the cosmological component in our data analysis we use 264 different theoretical 21-cm spectra presented by \citet{2017MNRAS.472.1915C}. In Fig. ~\ref{fig:eff} (top) we show a representative set of 25 input cosmological spectra in the  40--200 MHz band, from which the contribution of the CMB has been subtracted. To  demonstrate the effect of the SARAS~2 system, on the bottom panel of Fig. ~\ref{fig:eff} we show the same signals after they have been propagated through the system. The signals are attenuated when propagated through the system due to the total efficiency, with the loss increasing towards lower frequencies.

\subsection{Sensitivity Test}
\label{subsec:sens}

In this subsection our goal is to determine the optimal  frequency band $\Delta_{1,2}$, covering frequency range $\nu_1$ to $\nu_2$, and the order $N$ of $F(\nu)$  which provide  the best constraint on the particular signal template, $S_0(\nu)$. Using this information in Section \ref{subsec:data} we derive confidence with which each theoretical signal is  ruled out by the SARAS~2 data .  

We first perform a sensitivity test which, for each one of the 264 input templates $S_0(\nu)$ and given $\Delta_{1,2}$   and  $N$, determines whether or not the  signal can in principle be extracted from the data  considering the RMS thermal noise and the total efficiency of the system $\eta_t(\nu)$.  The test delineates the 2D  $\Delta_{1,2}$--$N$ parameter space in which the signal can be either detected or rejected with at least 1$\sigma$ confidence.

For given  $\Delta_{1,2}$ \&  $N$ we first generate 500 independent realizations of mock thermal noise with Gaussian statistics. The RMS thermal noise in any mock spectrum is matched to that in the data within the corresponding frequency band. We then add the propagated signal $S(\nu)$, in the frequency range $\nu_1$ to $\nu_2$, to each realization of the mock thermal noise creating 500 mock datasets. Each one of these datasets is then jointly fit with a model using linear least squares \cite[]{Press92numericalrecipes}:
\begin{equation}
M(\nu) = F(\nu) + a \times S(\nu),
\label{eq:joint}
\end{equation}
where $a$ is a scale factor for the signal.  The procedure returns best-fit values of the scale factor and coefficients of the polynomial, $c_i$, for each mock dataset separately. For each realization of the thermal noise, the fitting uncertainties in the polynomial coefficients, $\sigma_{c_i}$, and in the scale factor, $\sigma_a$, are computed as part of the modeling process from the covariance matrix.  We next perform joint fitting for all the 500 datasets and derive the mean, $\bar a$, and standard deviation,  $\sigma_{\bar a}$, for the scale factor across the ensemble of the mock datasets. 

For a detection, the extracted scale factor, $\bar a$, should be consistent with unity within the fitting uncertainties $\sigma_{\bar a}$. In other words, for each input signal $S_0(\nu)$ (and assuming the particular choice of $\Delta_{1,2}$ \& $N$) to be detected with more than $1\sigma$ confidence we require the following condition to be satisfied
\begin{equation}
0 < (\bar a - \sigma_{\bar a}) \leq 1 \leq (\bar a + \sigma_{\bar a}).
\label{Eq:cond}
\end{equation}
If this condition is not satisfied, we infer that the collected data (given its thermal RMS noise, $\Delta_{1,2}$ \& $N$)  is not sufficient to  detect  the particular $S_0(\nu)$ at $1\sigma$ level. 

This exercise ignores systematics that may leave residuals thus confusing detection of the 21-cm signal.  Therefore, it should be considered only as a feasibility test which helps to determine whether or not the RMS noise is sufficiently low for a detection with significance greater than $1 \sigma$. This sensitivity test affirms that if (i) the 21-cm signal is indeed present in the measurement data, and (2) there are no systematics limiting the decision, then the best fit results should yield $\bar a = 1$ with confidence exceeding $1\sigma$. 

Examination of the distribution of $\bar a$ for different $\Delta_{1,2}$ \& $N$  provides a 2D parameter space  ($\Delta_{1,2}$--$N$)  in which the condition above is satisfied. We use the allowed values of $\Delta_{1,2}$ \& $N$ in the next subsection to test each template against real data. If for a particular 21-cm signal the $\Delta_{1,2}$--$N$ parameter space is empty, this template is taken out of the ensemble and is not searched for. Therefore, the sensitivity test may be viewed as a preliminary filter that selects potentially good  candidate 21-cm signals which can be  detected/rejected using the collected data. We find that 27 models out of considered 264 cases pass the sensitivity test.

\subsection{Fitting the data}
\label{subsec:data}

\begin{figure}[htbp]
\begin{center}
\includegraphics[scale=0.3]{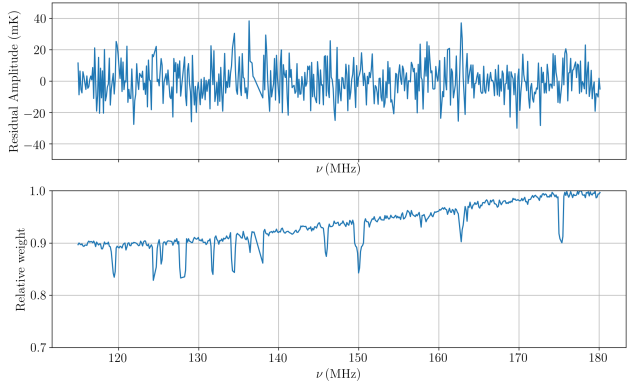}
\caption{The top panel shows representative data residuals at thermal noise levels.  The example shown was obtained after an $N =7^{th}$ order polynomial fit over the band of 110 to 180~MHz. The bottom panel shows the relative weights for the data in the frequency channels.}
\label{fig:2}    
\end{center}   
\end{figure}

We construct a set of models (Eq. \ref{eq:joint}) for each one of the 21-cm signals that pass the sensitivity test and for every combination of $\Delta_{1,2}$ \& $N$  from the allowed part of the parameter space. We fit every model to the real data using linear least squares. The  objective function defined as
\begin{equation}
\chi^2 = \sum_{\nu_1}^{\nu_2} w_{\nu_i}^2 (y_{\nu_i} - M(\nu_i))^2
\end{equation}
is minimized, where $y_{\nu_i}$ is the real data in the $i$th frequency channel and $M(\nu_i)$ is the model (Eq. \ref{eq:joint}).   $w_{\nu_i}$ are the relative weights for the data in each frequency channel $i$  based on the system temperature and effective integration times, which differ across the band depending on the RFI excision during the processing.  A representative data residual showing the thermal noise levels, and the corresponding relative weights across the band, are shown in Fig.~\ref{fig:2}.

In the fitting procedure to the SARAS~2 data, for each given theoretical 21-cm signal that passes the sensitivity test, the optimal $\Delta_{1,2}$ and $N$ are selected to be the combination for which the fit yields minimum uncertainty in the scale factor. The best fit scale factor is denoted as  $\tilde a$  with the  standard deviation,  $\sigma_{\tilde a}$,  given by the relevant diagonal term in the corresponding covariance matrix. In our analysis of all the plausible theoretical 21-cm signals in the atlas, the median value of the optimal $N$ is 4, and the associated frequency band is  110--180~MHz.  This is consistent with the fact that the foregrounds as well as internal systematics of SARAS 2 are indeed spectrally smooth and hence require only low-order polynomials for the modeling.  Typically, larger $N$ remove a greater part of the 21-cm signal; and,  therefore, return a larger uncertainty $\sigma_{\tilde a}$; while smaller $N$ are not sufficient to fit the foreground thus leaving behind larger residuals and increase the uncertainty $\sigma_{\tilde a}$.  

For each valid theoretical 21-cm signal we compute a standard score, $\zeta$, given by
\begin{equation}
\label{eq:zeta}
\zeta = \left| \frac{1-\tilde a}{\sigma_{\tilde a}} \right|.
\end{equation}
The value of $\zeta$ yields the confidence of the rejection in units of $\sigma_{\tilde a}$.  Based on this score we rule out any 21-cm signal with $\zeta > 1$, which ensures that the signal is inconsistent with the data at greater than $1\sigma$ confidence level. 

\begin{figure}[htbp]
\begin{center}
\includegraphics[scale=0.32]{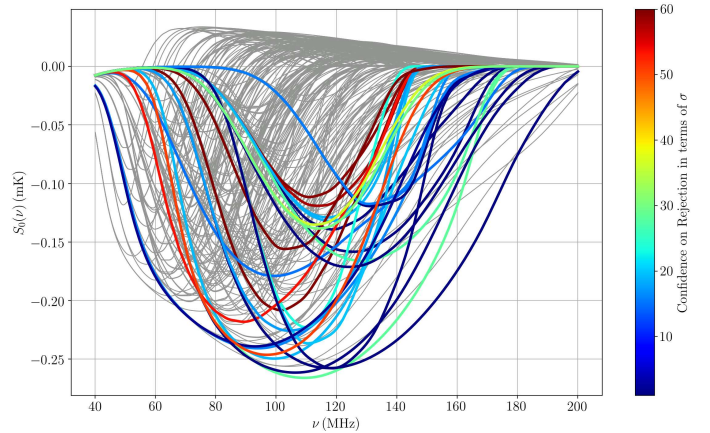}
\caption{The entire set of 264 theoretical models \citep{2017MNRAS.472.1915C}. The rejected signals are shown in color with each color corresponding to the rejection significance as indicated by the colorbar. The data does not have sensitivity for the signals shown in gray.}
\label{fig:3}    
\end{center}   
\end{figure}

For none of the considered theoretical models $\tilde a$ was found to be consistent with unity, which would indicate a detection. However, we find that for all the theoretical 21-cm signals which pass the sensitivity test, the condition for rejection is satisfied with confidence above 1$\sigma$; 25 templates have greater than  5$\sigma$ rejection significance. These cases are shown in colors in Figure \ref{fig:3} with each color representing the significance of rejection according to the colorbar.  Both the number of rejected cases and the significance of rejection are an improvement compared to our previous work  \citep{2041-8205-845-2-L12}. We note that very high values of rejection significance should be interpreted cautiously since the real data may include significant systematic noise with substantial non-Gaussianity.

\section{SARAS~2 constraints on 21-cm from CD/EoR}
\label{sec:cons}
In the parameter study conducted by \citet{2017MNRAS.472.1915C} the entire astrophysical parameter space, allowed by current observational and theoretical constrains, was sampled, and the 21-cm signals were derived for different combinations of the astrophysical parameters. In this study,  key astrophysical parameters were varied including the minimal circular velocity of star forming halos (starting from the minimal velocity of 4.2 km s$^{-1}$ characteristic for star formation via molecular cooling and up to 76.5 km s$^{-1}$), star formation efficiency (SFE) between 0.5\% and 50\%, spectral energy distribution (SED) of X-ray sources including hard and soft spectra \citep{2014Natur.506..197F}, X-ray efficiency compared to the low-redshift counterparts, mean free path (mfp) of ionizing radiation (cases with 5, 20 and 70  Mpc were considered), and the total optical depth, $\tau$.  The data collected by SARAS~2 is sufficient to rule out ~10\% of the considered theoretical models. 

The rejected models all share similar astrophysical properties: rapid reionization  in tandem with  either late X-ray heating due to very inefficient sources (13 cases) or  no heating at all (14 cases). In all these models the gas does not have enough time to heat up to the temperature of the CMB, and the 21-cm signal is seen in absorption throughout the EoR (colored lines in Fig. \ref{fig:3}). 

All the models ruled out by the SARAS~2 data share rapid reionization. We quantify this by the maximum rate of change of the brightness temperature of the 21-cm signal with respect to redshift, $(\frac{dS_0}{dz})_{\rm max}$. More than two-thirds of the rejected signals have $(\frac{dS_0}{dz})_{\rm max} > 86~\rm mK$ with the median value of the rejected set being 98~mK over $z \sim 10-6$ redshift interval. This is in contrast to the non-rejected signals where maximum $(\frac{dS_0}{dz})_{\rm max}$ is $86~\rm mK$ with median of 16~ mK over the same redshift range. Rapid reionization scenarios typically require one or more of the following: large mean free path of the ionizing photons, high star formation and ionizing efficiencies of the sources. All but 6 rejected cases have mfp of 70 Mpc; however, the values of SFE  and $\tau$ are unconstrained. The other 6 cases have mfp of 20 Mpc and high values of $\tau$. None of the rejected cases has mfp of 5 Mpc. 

Considering ``inefficient heating" models (sources with X-ray bolometric luminosity per star formation rate of up to 10\% of their low redshift counterparts) all the rejected cases share late star formation which only happens in massive halos with circular velocities above 35 km s$^{-1}$. In these cases the absorption trough is shifted into the SARAS~2 band, owing to the delayed build up of the Ly$\alpha$ background, making either detection or rejection easier. Majority of these cases have hard X-ray SED, while the value of SFE varies from model to model. 

The rejected astrophysical models with ``no heating'' have all possible values of circular velocities (from 4.2 to 76.5 km s$^{-1}$), SFE (from 0.5\% to 50\%) and values of $\tau$. Out of the 264 tested models, the only cases with ``no heating'' that were not ruled out have either very efficient star formation in light halos, and thus the absorption peak is shifted out of the SARAS~2 band, or have short mfp (5 Mpc) which implies more gradual reionization. 

A summary of astrophysical parameters for the rejected signals along with the significance of rejection is listed in Table~\ref{tab:param}.

\section{Caveats}
\label{sec:cav}

Experiments aiming to detect the global 21-cm from CD/EoR are difficult long-wavelength radiometer measurements, requiring a substantially wider dynamic range than what is typically necessary in most engineering applications at these frequencies.  Limitations may arise from unknowns in the internal systematics, antenna characteristics, ground emission, low level distributed RFI, and mode coupling of sky spatial structure into spectral measurement data owing to frequency dependent beams.

If the measurement equation describes the data to mK levels, including foregrounds and internal systematics, then a forward modeling approach is expected to be unbiased. This would apply even in the case of an excessive modeling of foreground+systematics with a higher than necessary order polynomial (which, however, would degrade the confidence in the derived results).  In an extreme case, if the model adopted for the foreground+systematics is also capable of fitting out the 21-cm template, the result would be completely ambiguous, with equal likelihoods for the presence and absence of the template.

Problems potentially arise when the  measurement equation or the adopted model is inadequate to describe the foreground+systematics, given the large dynamic range required for 21-cm signal detection. In this case, residual systematics can bias the results of the decision tests.  The adoption of an inadequate model may be inadvertent, particularly in the case where 21-cm signals are extracted via statistical analysis that aims to detect the signals in measurement data wherein the signal-to-noise ratio in individual channels are substantially below unity.  

Adopting an inadequate model would result in systematic residuals to the fit to foregrounds+systematics.  The least squares fit would attempt to maximize the correlation (or anti-correlation) of these residuals to the 21-cm template under consideration so that including a scale factor times the 21-cm template, the overall residuals would be a minimum.  Consequently, the unmodeled foreground+systematics might partially or wholly mimic the 21-cm signal---thus yielding a false positive---or partially or wholly cancel a true 21-cm signal in the data, thus yielding a false negative.  In these circumstances small fit residuals might suggest excellent fits with low formal statistical errors in the fitted scale factor $a$; however, the errors are obviously underestimates since the unmodeled systematics are not considered in the error computation. 

It is also necessary to consider cases where the true cosmological signal in the measurement data is substantially  different from the template used in modeling.  If the true cosmological signal is uncorrelated with the template, then the fit value of the scale factor $a$ will not be biased.  However, if there is partial correlation (or anti-correlation), then the fit would bias the $a$ parameter to be positive or negative depending on the correlation or anti-correlation respectively.  

The work presented herein has adopted polynomial models for the foreground+systematics.  Higher $N$ would obviously fit this term  better; however, it would also increasingly subsume the cosmological signal and, hence, reduce the confidence in either detection or rejection.  Future effort is directed to improve the modeling of foregrounds+systematics and avoid fitting out of a significant fraction of 21-cm signals.  The design of SARAS~2, which aims to constrain the systematics to be maximally smooth \cite[]{2017arXiv171001101S} is along the lines of this approach.  

\section{Conclusion}
\label{sec:conclusion}
In this work we have analyzed the first light data from SARAS~2 using a frequentist approach and forward modeling. The revised analysis has led to the rejection of 27 plausible 21-cm signals out of 264 tested models. In 25 cases the confidence on rejection is above 5$\sigma$. All the rejected signals lie in the regime of either late or non-existent heating by the first population of X-ray sources which creates a deep absorption trough in the 21-cm signal observed against the CMB. In addition, in all the rejected models reionization happens fast owing to the assumed long mean free path of the ionizing photons as well as efficient star formation and ionization. We leave robust estimation of the rejected parameter space to future work. 

\section*{Acknowledgement}

We thank staff at Gauribidanur Field Station for assistance with
system tests and measurements, and the Mechanical 
and Electronics Engineering Groups at Raman Research Institute for
building and assembling SARAS~2. Santosh Harish and Divya Jayasankar
implemented real-time software and monitoring. Logistics and technical
support for observations was provided by Indian Astronomical
Observatory, Leh operated by Indian Institute of Astrophysics, and
Timbaktu Collective, India. For R.B.\ and A.C.\, this
project/publication was made possible through the support of a grant
from the John Templeton Foundation. The opinions expressed in this
publication are those of the author(s) and do not necessarily reflect
the views of the John Templeton Foundation. This research was also supported (for R.B. and A.C.) by the ISF-NSFC
joint research program (grant No. 2580/17).

\begin{table*}
\centering
\begin{threeparttable}
\caption{Parameters of the 21-cm signals rejected by the SARAS~2 data.}
\label{tab:param}
\begin{tabular}{||cccccccc||}
\hline
\toprule
$f_*$ & $V_c$ (km/s) & $f_X$& SED & $\tau$ & $R_{\rm mfp}$ & $(\frac{dS_0}{dz})_{\rm max}$ & Significance of rejection \\ 
\midrule
\hline\hline
0.050 & 35.50 & 0    & Hard & 0.06  & 70 & 118.22 & 1.06  \\
\hline
0.500 & 4.20  & 0    & Hard & 0.082 & 20 & 86.14  & 2.02  \\
\hline
0.015 & 76.50 & 0.1  & Soft & 0.066 & 70 & 159.21 & 12.10 \\
\hline
0.005 & 4.20  & 0    & Hard & 0.082 & 70 & 74.32  & 15.04 \\
\hline
0.015 & 76.50 & 0.1  & MQ & 0.066 & 70 & 146.62 & 15.54 \\
\hline
0.500 &  4.20  & 0    & Hard & 0.082 & 70 & 164.83 & 16.15 \\
\hline
0.050 &  35.50 & 0    & Hard & 0.083 & 70 & 164.71 & 17.57 \\
\hline
0.005 &  35.50 & 0.1  & Hard & 0.082 & 70 & 110.48 & 18.60 \\
\hline
0.005 & 35.50 & 0    & Hard & 0.082 & 70 & 118.34 & 18.91 \\
\hline
0.500 & 35.50 & 0    & Hard & 0.082 & 70 & 172.80 & 19.05 \\
\hline
0.500 & 76.50 & 0.1  & Hard & 0.066 & 70 & 97.17  & 19.71 \\
\hline
0.500 & 76.50 & 0.1  & MQ & 0.066 & 70 & 169.85 & 20.18 \\
\hline
0.005 & 35.50 & 0.1  & Hard & 0.066 & 70 & 104.18 & 22.16 \\
\hline
0.500 &  76.50 & 0.1  & Hard & 0.082 & 70 & 128.98 & 23.52 \\
\hline
0.005 & 35.50 & 0    & Hard & 0.066 & 70 & 131.70 & 27.95 \\
\hline
0.500 & 35.50 & 0    & Hard & 0.066 & 70 & 172.59 & 28.22 \\
\hline
0.005 & 35.50 & 0    & Hard & 0.082 & 20 & 67.40  & 29.99 \\
\hline
0.005 & 35.50 & 0.1  & Hard & 0.082 & 20 & 64.36  & 36.39 \\
\hline
0.005 & 35.50 & 0    & Hard & 0.066 & 20 & 74.87  & 36.71 \\
\hline
0.500 & 35.50 & 0    & Hard & 0.082 & 20 & 94.45  & 50.65 \\
\hline
0.050 & 16.50 & 0    & Hard & 0.096 & 70 & 94.92  & 53.67 \\
\hline
0.005 & 35.50 & 0.1  & Soft & 0.082 & 70 & 88.33  & 57.81 \\
\hline
0.005 & 35.50 & 0.1  & Soft & 0.066 & 70 & 55.45  & 58.83 \\
\hline
0.005 & 35.50 & 1    & Hard & 0.082 & 70 & 69.71  & 59.35 \\
\hline
0.050 & 35.50 & 0.1  & Hard & 0.082 & 70 & 94.26  & 63.02 \\
\hline
0.500 & 35.50 & 0    & Hard & 0.066 & 20 & 97.69  & 67.94 \\
\hline
0.015 & 35.50 & 0.16 & Hard+MQ & 0.082 & 70 & 75.28  & 68.17 \\
\hline
\bottomrule
\end{tabular}
\begin{tablenotes}
     \small
      \item $f_*$ denotes the star formation efficiency, $V_c$ represents minimum virial circular velocity for star formation, $f_X$ is the efficiency of the X-ray sources, SED refers to spectral energy distribution of X-ray sources. The SEDs considered are of hard and soft X-ray sources along with that of mini-quasars (MQ). $\tau$ is CMB optical depth, $R_{\rm mfp}$ denotes the mean free path of ionizing photons, $(\frac{dS_0}{dz})_{\rm max}$ is the maximum rate of change of brightness temperature of the signal with respect to redshift. Significance of rejection is computed as given in Eq.~\ref{eq:zeta}. A detailed description of most of these parameters is given in \cite{2017MNRAS.472.1915C}.
    \end{tablenotes}
  \end{threeparttable}
\end{table*}

\end{document}